# Thick-target yields of radioactive targets deduced from inverse kinematics

M. Aikawa[a,1], S. Ebata[b], S. Imai[b]

[a] Faculty of Science, Hokkaido University, Sapporo 060-0810, Japan
[b] Meme Media Laboratory, Hokkaido University, Sapporo 060-8628, Japan

**Abstract**

The thick-target yield (TTY) is a macroscopic quantity reflected by nuclear reactions and matter properties of targets. In order to evaluate TTYs on radioactive targets, we suggest a conversion method from inverse kinematics corresponding to the reaction of radioactive beams on stable targets. The method to deduce the TTY is theoretically derived from inverse kinematics. We apply the method to the $^{nat}Cu(^{12}C,X)^{24}Na$ reaction to confirm availability. In addition, it is applied to the $^{137}Cs + ^{12}C$ reaction as an example of a radioactive system and discussed a conversion coefficient of a TTY measurement.

Nuclear data of radioactive isotopes (RI) are important for astrophysics, nuclear physics and nuclear engineering. However the experiments with radioactive targets are limited due to the high radioactivity. A method to obtain the data is to utilize the experimental values measured in inverse kinematics. RI beams are available in accelerators to obtain nuclear data. The beam has been applied to obtain cross sections directly [1] and deduced from the thick-target method [2].The experiments with heavy RI are also realized by the recent progress in radioactive beams [3].

In addition to cross section data, an essential quantity is the thick-target yield (TTY) [4]. The TTYs can be theoretically defined and estimated from cross sections and stopping powers [5].

---

There are previous researches using RI beams to obtain cross section and TTY data [6,7,8]. However, the TTYs on radioactive targets are hard to measure directly in accelerators since the preparation of radioactive lumps as a target is very difficult due to the high radioactivity. We suggest a new method to estimate TTYs using inverse kinematics even though the cross section and TTY data of individual reaction channels are not provided.

Reaction probability $Y$ is defined as

$$Y = \frac{N_r}{N_i}, \tag{1}$$

where $N_r$ and $N_i$ are the numbers of reacted and incident particles. In the case of a thick target, the $Y$ is equivalent to TTY of the reaction and the differential reaction probability $dY$ at an infinitesimal length in the target $dx$ [cm] can be described as:

$$dY = \sigma \frac{\rho N_A}{A_T} dx, \tag{2}$$

where the cross section $\sigma$ [cm$^2$], the Avogadro constant $N_A$ [mol$^{-1}$], the mass number of the target $A_T$ [g·mol$^{-1}$] and the density $\rho$ [g·cm$^{-3}$]. We can obtain the TTY by integrating the thickness of the target and introducing a stopping power $S(E) = -\frac{dE}{d(\rho x)}$ [MeV·g$^{-1}$·cm$^2$]:

$$Y(E_{in}) = \frac{N_A}{A_T} \int_0^{E_{in}} \sigma(E) \frac{1}{S(E)} dE, \tag{3}$$

at the incident energy in a laboratory system $E_{in}$ [MeV]. The lower limit of energy integration of Eq. (3) corresponds to zero, i.e. the incident particles are stopped inside the target. We assume here that the incident beam current is almost constant [4,5]. The integration variable and the upper limit can be converted into energy per nucleon $\varepsilon = \frac{E}{A_P}$ with the mass number of the projectile $A_P$ and its incident energy per nucleon $\varepsilon_{in}$. Using $S(\varepsilon) = -\frac{A_P d\varepsilon}{d(\rho x)}$, the TTY is rewritten as:

$$Y(\varepsilon_{in}) = \frac{N_A A_P}{A_T} \int_0^{\varepsilon_{in}} \sigma(\varepsilon) \frac{1}{S(\varepsilon)} d\varepsilon, \tag{4}$$

which leads to:

$$\frac{dY(\varepsilon)}{d\varepsilon} = \frac{N_A A_P}{A_T} \sigma(\varepsilon) \frac{1}{S(\varepsilon)}. \tag{5}$$

We note that the Eqs. (4) and (5) are available under the condition that all the projectiles stop inside the target, namely, the incident energy $\varepsilon_{in}$ should be less than the minimum energy to pass through the target.

In this paper, we define the projectile $P$ induced reaction on a target $T$ as "forward" and its inverse kinematics as "inverse". Those TTYs which are evaluated from Eq. (4) are denoted as $Y_{\text{for}}$ and $Y_{\text{inv}}$. We consider the relation between $Y_{\text{for}}$ and $Y_{\text{inv}}$ using the inverse kinematics and the common cross section in both reactions. Hence, the ratio $R(\varepsilon)$ between the differential yields at the same $\varepsilon$ in both reactions can be expressed as

$$R(\varepsilon) \equiv \frac{dY_{\text{for}}}{dY_{\text{inv}}} = \frac{A_P^2}{A_T^2} \frac{S_{\text{inv}}(\varepsilon)}{S_{\text{for}}(\varepsilon)}. \qquad (6)$$

This relation suggests that we can evaluate the $Y_{\text{for}}(\varepsilon)$ without the direct experiment of the forward kinematics, in cases where the $Y_{\text{inv}}(\varepsilon)$ and the $R(\varepsilon)$ are obtained. The heavier projectile is more easily stopped inside the target since the stopping power is proportional to the square of the projectile atomic number $Z_P^2$ and the target atomic number $Z_T$ in the high energy region. This also implies an advantage of the method because the target of the light elements can be thin and prepared more easily.

We show the examples of the $^{\text{nat}}$Cu($^{12}$C,X)$^{24}$Na [9] and $^{137}$Cs + $^{12}$C reactions to confirm the procedure, as well as its availability. The method is initially applied to find the relation between the $^{\text{nat}}$Cu($^{12}$C,X)$^{24}$Na forward and $^{12}$C($^{63,65}$Cu,X)$^{24}$Na inverse kinematics reactions. In order to calculate the $Y_{\text{for}}(\varepsilon)$ and $Y_{\text{inv}}(\varepsilon)$ from Eq. (4), the cross section $\sigma(\varepsilon)$ and stopping power $S(\varepsilon)$ are necessary. The $\sigma(\varepsilon)$ as a function of $\varepsilon$ is prepared by the spline fitting of experimental data [9], while the $S(\varepsilon)$ is computed using the SRIM2008 code [10]. The $\sigma(\varepsilon)$ and $S(\varepsilon)$ are shown in Figs. (1) and (2)(a). The $Y_{\text{for}}(\varepsilon)$ with $A_T(^{\text{nat}}$Cu$) = 63.546$ at $\varepsilon = 40$ and $100$ [MeV/nucleon] are evaluated by Eq. (4) and the results are $Y_{\text{for}}(40) = 0.91 \times 10^{-5}$ and $Y_{\text{for}}(100) = 0.114 \times 10^{-3}$ (Table 1). We can also obtain the TTY of inverse kinematics, $Y_{\text{inv}}(40) = 0.86 \times 10^{-5}$ and $Y_{\text{inv}}(100) = 0.103 \times 10^{-3}$, in the same manner as the $Y_{\text{for}}$. On the other hand, $Y_{\text{for}}(\varepsilon)$ can be calculated from $R(\varepsilon)dY_{\text{inv}}(\varepsilon)$ in Eq. (6) and vice versa. Here, we note that the stopping powers $S(\varepsilon)$ shown in Fig. (2)(a) are largely different from each other, but the $R(\varepsilon)$ converges to a constant value at the high energy over 50 MeV/nucleon shown in Fig. (2)(b). The cross section of $^{\text{nat}}$Cu($^{12}$C,X)$^{24}$Na is negligibly small in the low energy region since such fragmentation reaction requires a large amount of energy. This simple behavior of $R(\varepsilon)$ in Eq. (6) and the small $\sigma(\varepsilon)$ allows us to utilize a more convenient conversion method as:

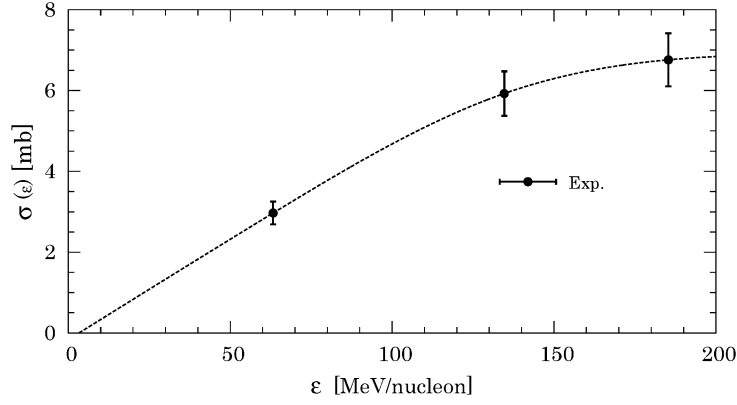

Fig. 1: Cross section with respect to $\varepsilon = \frac{E}{A_P}$ of the $^{nat}$Cu($^{12}$C,X)$^{24}$Na taken from experimental data [9]. Dashed-line is the spline fitting.

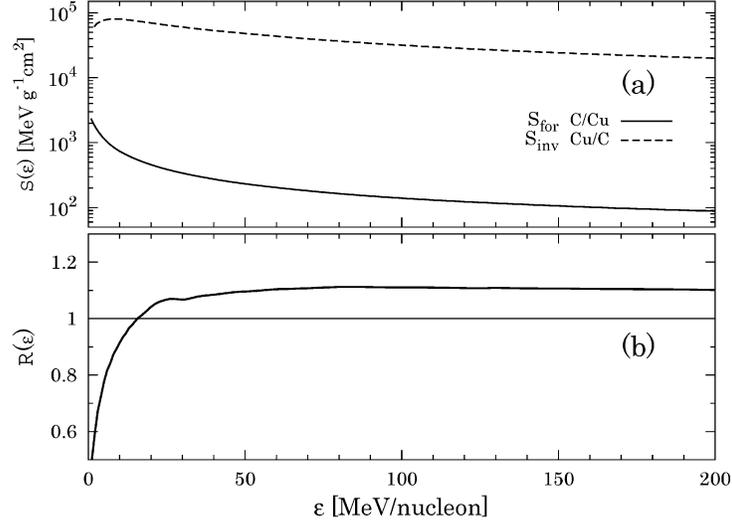

Fig. 2: (a) Stopping powers $S(\varepsilon)$ and (b) ratio $R(\varepsilon)$ of the Cu + C system by SRIM2008 [10].

Table 1: $Y_{\text{for}}(\varepsilon)$ at 40 and 100 MeV/nucleon derived from Eqs. (4) and (7).

|        | $Y_{\text{for}}(40)$ | $Y_{\text{for}}(100)$ |
|--------|----------------------|------------------------|
| Eq. (4) | $0.91 \times 10^{-5}$ | $0.114 \times 10^{-3}$ |
| Eq. (7) | $0.94 \times 10^{-5}$ | $0.113 \times 10^{-3}$ |

$$Y_{\text{for}}(\varepsilon) \cong \tilde{R} Y_{\text{inv}}(\varepsilon), \tag{7}$$

where $\tilde{R}$ is a constant value of $R(\varepsilon)$ at the high energy over 50 MeV/nucleon in this case. We can estimate $Y_{\text{for}}$ from $Y_{\text{inv}}$ and $\tilde{R}_{\text{Cu/C}} = 1.1$ and obtain $Y_{\text{for}}(40) = 0.94 \times 10^{-5}$ and $Y_{\text{for}}(100) = 0.113 \times 10^{-3}$. These are in good agreement with values derived from Eq. (4) as shown in Table 1. This conversion method is practically justified using the SRIM2008 code and a negligible cross section at the low energy. We emphasize here that the experimental value of $Y_{\text{inv}}$ is also available to estimate $Y_{\text{for}}$ from Eq. (7) without the cross section. If the TTYs of $^{\text{nat}}\text{Cu}(^{12}\text{C,X})^{24}\text{Na}$ and $^{12}\text{C}(^{65}\text{Cu,X})^{24}\text{Na}$ can be measured experimentally, we can prove the validity of the method.

Next, we consider the inclusive reaction of $^{137}\text{Cs}$ induced by $^{12}\text{C}$ which consists of all channels except for the $^{137}\text{Cs}(^{12}\text{C,X})^{137}\text{Cs}$ as the forward kinematics. Its inverse kinematics reaction has the $^{12}\text{C}$ target and the radioactive $^{137}\text{Cs}$ projectile. The $Y^{\text{incl.}}$ of the inclusive reaction bombarding the thick-target is described as:

$$Y^{\text{incl.}} = \frac{N_r}{N_i} = \frac{N_i - N_u}{N_i}, \tag{8}$$

where $N_r$, $N_i$, and $N_u$ are the number of the inclusively reacted, incident, and un-changed particles, respectively. $N_i$ is a countable number experimentally, and $N_u$ can also be observed through detection of the specific gamma-rays if the projectile is radioactive and can be identified by gamma decay modes. The $Y^{\text{incl.}}$ of a radioactive projectile is therefore obtainable.

In order to evaluate the $Y_{\text{for}}^{\text{incl.}}$ for the inclusive reaction of $^{137}\text{Cs}$, we calculate the $R(\varepsilon)$ in Eq. (6) using the SRIM2008 code. The $R(\varepsilon)$ of the reaction of $^{137}\text{Cs}$ beam on the $^{12}\text{C}$ target (Cs/C) is shown in Fig. (3). The plateau and convergence of $R(\varepsilon)$ can be seen in the high energy region as is the case with the previous example. We can find that the Cs/C system is not a special case since the ratios of two other samples, zirconium (Zr/C) and uranium (U/C), show similar tendencies.

The value at the plateau is the maximum $R(\varepsilon)$ of the system ($\cong 1.05$). On the contrary, the minimum value can be estimated from Coulomb barrier which defines the lowest energy of reactions. The peak energy of Coulomb barrier can be roughly estimated as:

$$U_C = \frac{Z_T Z_P e^2}{1.2\left(A_T^{1/3} + A_P^{1/3}\right)}. \tag{9}$$

In the $^{137}\text{Cs} + ^{12}\text{C}$ system, $U_C \cong 4.82$ MeV/nucleon and $R(\varepsilon) \cong 0.51$ at minimum are obtained.

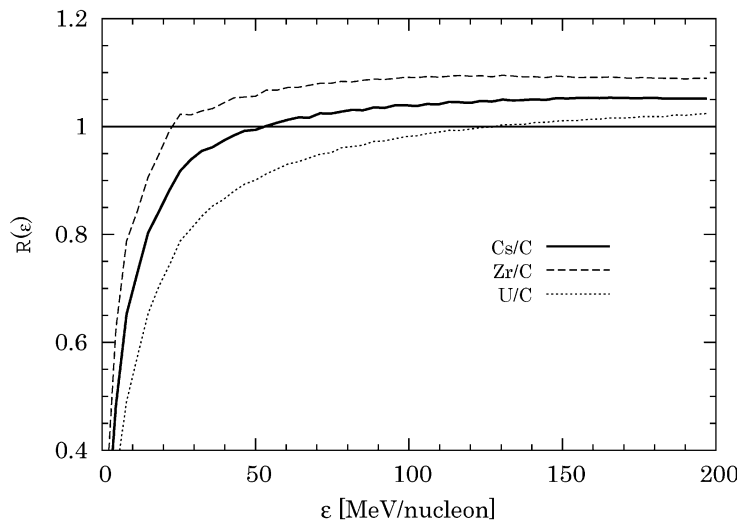

Fig. 3: Ratio of differential yields between cesium and carbon (Cs/C) in comparison with two other samples, zirconium (Zr/C) and uranium (U/C).

The TTY is a key quantity in addition to the cross section. While there is a program to measure cross sections on minor actinides [11], the direct measurement remains difficult due to the high radioactivity. We suggest a new method to estimate the TTY using inverse kinematics. As an example, we show that the TTY of the carbon-induced reaction on the copper lump is estimated from the copper-induced reaction on the carbon target. We derived the TTYs from Eqs. (4) and (7), respectively, and found a good agreement of them. We can also estimate a conversion coefficient $R(\varepsilon)$ of inclusive reaction of $^{137}$Cs induced by $^{12}$C from the $Y_{\text{inv}}^{\text{incl.}}$ within a range between 0.51 and 1.05. This method is highly effective to a radioactive system, and is applicable to transmutation reactions to reduce long-lived fission products. More precise analysis using Monte Carlo simulation codes and other cases with different projectiles and targets will be discussed in the forthcoming papers.

**Acknowledgement**

The authors are grateful to Dr. H. Sakurai, Dr. H. Otsu in RIKEN, Dr. K. Ogata in Osaka University and Dr. N. Otuka in IAEA for valuable discussion. The authors also thank all members in Hokkaido University Nuclear Reaction Data Centre (JCPRG) for their supports.